\documentclass[preprint,prb,showpacs]{revtex4}

\usepackage{graphicx}
\usepackage{dcolumn}
\usepackage{bm}
\usepackage{natbib}

\begin{document}

\title{\boldmath Extended Drude model analysis of superconducting optical spectra of correlated electron systems \unboldmath}

\author{Jungseek Hwang}
\email{jungseek@skku.edu}
\affiliation{Department of Physics, Sungkyunkwan University, Suwon, Gyeonggi-do 16419, Republic of Korea}

\date{\today}

\begin{abstract}
Correlation information in strongly correlated electron systems can be obtained using an extended Drude model. An interesting method related to the extended Drude model analysis of superconducting optical data was proposed recently, and it has attracted attention from researchers. This method aims to extract the optical self-energy of quasiparticles (or residual unpaired electrons) from measured optical data in the superconducting state. However, this residual optical self-energy is a partial optical self-energy. The interpretation and significance of this partial optical self-energy is unclear. We investigate this method using a reverse process with simple electron–boson spectral density functions. With our obtained results, we conclude that the residual (or partial) optical self-energy is difficult to interpret because it contains unphysical features, in particular, a negative optical effective mass. The present study clarifies the extended Drude analysis method for superconducting optical data.
\end{abstract}

\maketitle

\section{Introduction}

Charge carriers in strongly correlated electron systems can be described by an extended Drude model, which is a single-band model approach. In this model, a new physical quantity appears, the memory function \cite{gotze:1972} or the optical self-energy\cite{hwang:2004}; this new quantity is a measure of the strength of correlation and is closely related to the well-known quasiparticle self-energy\cite{carbotte:2005,hwang:2007b}. Furthermore, because optical processes (or transitions) occur between filled and empty states, i.e., two states (or two particles) are involved in the process; therefore, the optical self-energy is also called a two-particle self-energy, which is mathematically more complicated than its quasiparticle counterpart. In a strongly correlated regime, i.e., in an extreme case, the single band splits into two distinct bands (upper and lower Hubbard bands)\cite{kotlier:2004}. However, in an intermediately correlated regime, the single band is decomposed into coherent and incoherent components. The extended Drude formalism can be applied to various strongly correlated electron systems in the intermediate regime for both the normal and superconducting states. Thus far, many important and intriguing results\cite{hwang:2004,webb:1986,sulewski:1988,thomas:1988,awasthi:1993,puchkov:1996,carbotte:1999,dordevic:2001,tran:2002,dordevic:2005,heumen:2009,hwang:2016} have been obtained from the measured optical spectra of various correlated electron systems using the extended Drude model approach.

One can extract the electron-boson spectral density function (or Eliashberg function) from a measured reflectance spectrum using a well-known process\cite{collins:1989,carbotte:1990,schachinger:2000,hwang:2006,hwang:2007}, called a normal process, consisting of several analysis steps: the Kramers–Kronig relation, extended Drude model, generalized Allen's integral equations, and numerical processes to solve the Allen's integral equations\cite{schachinger:2006,hwang:2015a}. One should be able to calculate a reflectance spectrum and other optical response functions from a known electron-boson spectral density function using a reverse process\cite{hwang:2015a}, which consists of the same analysis steps in reverse order of the normal process. The reverse process\cite{hwang:2015a} can be used for various purposes, such as to study how some characteristic features in the electron-boson density function appear in other optical response functions including a reflectance spectrum, and to extract the governing electron-boson spectral density function\cite{hwang:2016} of multiband correlated electron systems (e.g., iron pnictides) by simulating the measured optical data.

In this study, we address a recently proposed method by Dordevic {\it et al.}\cite{dordevic:2014} regarding the extended Drude model analysis of optical data in the superconducting state. Dordevic {\it et al.} stated that the application of the extended Drude model to data in the superconducting state is has many limitations. Furthermore, they stated that the extended Drude model (or single-band approach) is violated in the superconducting state because normal fluids and superfluids coexist in the superconducting state. This proposed method may allow access to the intrinsic properties of quasiparticles (or residual unpaired electrons) in the superconducting state. In addition, they introduced a new self-energy for the quasiparticles in the superconducting state, which we name the residual optical self-energy in this paper. It is noteworthy that the residual optical self-energy is a partial optical self-energy in the extended Drude formalism. We investigated this method in detail with a correlated electron formalism using the inverse process with well-known typical electron-boson spectral density functions\cite{hwang:2015a} in the $d$-wave superconducting material phase. These typical electron–boson spectral density functions are well established in terms of doping- and temperature-dependent evolutions\cite{carbotte:2011,hwang:2007,hwang:2016a,hwang:2018}. Furthermore, the generic phase diagram of hole-doped cuprates can be described well with these typical models\cite{hwang:2018}. From our calculation results, we conclude that the residual (or partial) optical self-energy suffers from serious problems, because it exhibits unphysical features. Meanwhile, the full SC optical self-energy is a reliable optical quantity as it does not exhibit these problems. In addition, we applied this method to the existing measured optical data of two differently doped (i.e., optimally doped and overdoped) Bi$_2$Sr$_2$CaCu$_2$O$_{8+\delta}$ (Bi2212) cuprate systems in the superconducting phase. The resulting residual (or partial) self-energies of the measured Bi2212 spectra show the same serious problems as those in the theoretical calculations, implying that the proposed partial optical self-energy cannot be applied to measured optical data in the SC phase, even though the idea is intriguing.

\section{Reverse process formalism}

In this section, we briefly describe the reverse process\cite{hwang:2015a} for the two (normal and $d$-wave superconducting) material phases. Note that both the normal and superconducting phases are of correlated electron systems. The interaction between electrons can occur by exchanging force-mediating bosons, and it can be described with the electron-boson spectral density function, $\alpha^2 B(\omega)$, where $\alpha$ and $B(\omega)$ are the coupling constant between an electron and the boson and the boson spectral density, respectively. With an input model $\alpha^2 B(\omega)$ one can obtain the imaginary part of the optical self-energy [$\tilde{\Sigma}^{op}(\omega) \equiv \Sigma_1^{op}(\omega) + i\Sigma_2^{op}(\omega)$] using the generalized Allen's formalism\cite{allen:1971} as
\begin{equation}\label{eq1}
-2\Sigma_2^{op}(\omega) =  \int^{\infty}_{0} d\Omega \: \alpha^2 B(\Omega) \: K(\omega, \Omega)  + \Gamma^{op}_{imp}(\omega),
\end{equation}
where $K(\omega, \Omega)$ and $\Gamma^{op}_{imp}(\omega)$ are the kernel of the generalized Allen's integral equation and the optical impurity scattering rate, respectively. For simplicity without losing the generality one may consider cases only at $T=$ 0. In such cases, the kernels for normal and $d$-wave superconducting\cite{allen:1971,schachinger:2006} states can be written as
\begin{eqnarray}\label{eq2}
K(\omega, \Omega)&=& 2\pi \Big{(}1-\frac{\Omega}{\omega}\Big{)} \Theta(\omega-\Omega) \:  \:\: (\mbox{for normal state})\nonumber \\
  &=&2\pi \Big{(}1-\frac{\Omega}{\omega}\Big{)} \Big{\langle}\Theta(\omega-2\Delta(\theta)-\Omega) \nonumber \\
  &\times& E\Big{(} \sqrt{1-\frac{4 \Delta(\theta)^2}{(\omega-\Omega)^2}} \Big{)} \Big{\rangle}_{\theta} \:\:(\mbox{for $d$-wave SC}),
\end{eqnarray}
where $\Theta(x)$ is the Heaviside step function, $E(x)$ is the complete elliptic integral of the second kind, and $\langle\cdots\rangle_{\theta}$ represents the angular average over a range from 0 to $\pi/4$. $\Delta(\theta)$ [=$\Delta_0 \cos(2\theta)$] is the $d$-wave superconducting (SC) gap, where $\Delta_0$ is the maximum SC gap.

The optical impurity scattering rate can be written as\cite{allen:1971,schachinger:2006}
\begin{eqnarray}\label{eq3}
\Gamma^{op}_{imp}(\omega) &=& \!\!\Gamma_{imp} \:\: (\mbox{for normal state})\nonumber \\
        &=& \!\!\Gamma_{imp} \Big{\langle} E\Big{(} \sqrt{1\!-\!\!\frac{4 \Delta(\theta)^2}{\omega^2}} \Big{)} \Big{\rangle}_{\theta} (\mbox{for $d$-wave SC}),
\end{eqnarray}
where $\Gamma_{imp}$ is a constant impurity scattering rate. It should be noted that the impurity scattering rate in the SC state is strongly dependent on frequency near twice of the $d$-wave superconducting gap, 2$\Delta(\theta)$.

Once one has the imaginary part of the optical self-energy in a wide enough spectral range one can get the corresponding real part using a Kramers-Kronig relation as
\begin{equation}\label{eq4}
-2\Sigma_1^{op}(\omega) = -\frac{2\omega}{\pi}P\int^{\infty}_0 d\nu  \: \frac{[-2\Sigma_2^{op}(\nu)]}{\nu^2-\omega^2},
\end{equation}
where $P$ presents the principle part of the improper integral. The mass enhancement function [$m_{op}^*(\omega)/m_b$] can be calculated from the real part using a relation between the two quantities as $m_{op}^*(\omega)/m_b = -2\Sigma_1^{op}(\omega)/\omega + 1$, where $m_{op}^*(\omega)$ and $m_b$ are the optical effective mass and the band mass, respectively. It is worth noting that for convergence of the Kramers-Kronig (KK) relations between the real and imaginary parts of a complex function the function generally needs to be square integrable. However, the optical self-energy is not square integrable. Still the Kramers-Kronig relations can be valid if $\lim_{\omega \rightarrow \infty}|-2\tilde{\Sigma}^{op}(\omega)/\omega|$ is zero. Indeed, the magnitude of the optical self-energy divided by the frequency approaches to zero as $\omega \rightarrow \infty$ since $\lim_{\omega \rightarrow \infty}[-2\Sigma_2^{op}(\omega)] = 2\pi \times$ the area under $\alpha^2B(\omega)$\cite{hwang:2008a} and $\lim_{\omega \rightarrow \infty}[-2\Sigma_1^{op}(\omega)/\omega] \equiv \lim_{\omega \rightarrow \infty}[m_{op}^*(\omega)/m_ b - 1] =$ 0 (see Fig. 5 of Ref. [27]). This requirement is weaker than the square integrable one; in this case, the KK relations are known as KK relations with one subtraction\cite{nussenzveig:1972}.

Finally, using the obtained complex optical self-energy one can calculate the optical conductivity [$\tilde{\sigma}(\omega) \equiv \sigma_1(\omega)+i\sigma_2(\omega)$] through the extended Drude formalism\cite{gotze:1972,hwang:2004} as
\begin{equation}\label{eq5}
\tilde{\sigma}(\omega)=\frac{i}{4\pi}\frac{\Omega_p^2}{\omega+[-2\tilde{\Sigma}^{op}(\omega)]},
\end{equation}
where $\Omega_p$ is the plasma frequency, which is directly associated with the charge carrier density ($n$) in the mode of interest, i.e., $\Omega_p^2 = 4 \pi n e^2/ m_b$, where $e$ is the elementary charge. Further, it should be noted that the extended Drude model is a single-band approach; here we call the intraband transition in the single band as "the extended Drude mode", which usually consists of two (coherent and incoherent) components, as mentioned in the introduction. Most of the spectral weight of the extended Drude mode is usually located in the low-energy region close to the Fermi energy. However, the full extended Drude mode should be extended to infinity as per the simple Drude mode. If a theoretical system contains only a single extended Drude mode, the range of validity of the extended Drude model must be extended from zero frequency to infinity. In this case, the real and imaginary parts of the optical self-energy defined by the extended Drude model should form a Kramers–Kronig pair; one can obtain the real part of the optical self-energy from the known imaginary part using Eq. (4). In contrast, because a real (material) system usually contains many modes (or oscillators) including the extended Drude mode (located in the lowest frequency region), the range of validity of the extended Drude model is extended below the first interband transition mode, i.e., the contributions from all interband transition modes located in the high-energy region must be excluded. In this case, the real and imaginary parts of the optical self-energy defined by the extended Drude model generally cannot form a Kramers–Kronig pair because the optical conductivity consists of many modes including the extended Drude mode.

In the SC state the optical conductivity [$\tilde{\sigma}^{SC}(\omega) = \sigma_1^{SC}(\omega) +i \sigma_2^{SC}(\omega)$] can be divided into two components as
\begin{eqnarray}\label{eq8}
\sigma_1^{SC}(\omega)&=& \frac{\Omega_{sp}^2}{8}\delta(\omega) + \sigma_1^{res}(\omega) \nonumber \\
\sigma_2^{SC}(\omega)&=& \frac{\Omega_{sp}^2}{4\pi \:\omega} + \sigma_2^{res}(\omega),
\end{eqnarray}
where $\Omega_{sp}$ is the superfluid plasma frequency, which should be smaller than the plasma frequency ($\Omega_p$), and $\delta(\omega)$ is the Dirac delta function. $\sigma_1^{res}(\omega)$ and $\sigma_2^{res}(\omega)$ are, respectively, the real and imaginary parts of the residual optical conductivity [$\tilde{\sigma}^{res}(\omega)$], which form a Kramers-Kronig (KK) pair. It should be noted that, in a finite frequency region (i.e. $\omega >$ 0), $\sigma_1^{SC}(\omega)$ and $\sigma_1^{res}(\omega)$ are equal to each other but they are different at $\omega =$ 0, i.e, $\sigma_1^{SC}(0) = \Omega_{sp}^2 \delta(0)/8$ and $\sigma_1^{res}(0) =$ 0. Therefore, only $\sigma_1^{res}(\omega)$ part of the full $\sigma_1^{SC}(\omega)$ can be seen in a measured finite frequency region and $\sigma_2^{res}(\omega)$ is not directly accessible. However, $\sigma_2^{res}(\omega)$ can be obtained from $\sigma_1^{res}(\omega)$ using the Kramers-Kronig relation\cite{wooten} since they form a KK pair. Once $\sigma_2^{res}(\omega)$ is obtained the superfluid plasma frequency can be estimated using the second equation of Eq. (6) as
\begin{equation}\label{eq9}
\Omega_{sp}^2 = 4 \pi \omega [\sigma_2^{SC}(\omega) - \sigma_2^{res}(\omega)].
\end{equation}
The equation above can be rewritten as
\begin{equation}\label{}
\Omega_{sp}^2 = \omega^2[\epsilon_H - \epsilon_1^{SC}(\omega)] - 4\pi \omega \sigma_2^{res}(\omega),
\end{equation}
where $\epsilon_H$ is the high-frequency background dielectric constant and $\epsilon_1^{SC}(\omega)$ is the real part of the full SC dielectric function, which is related to the full SC optical conductivity as $\tilde{\epsilon}^{SC}(\omega) = \epsilon_H+i 4 \pi \tilde{\sigma}^{SC}(\omega)/\omega$.
Since $\epsilon_H$ is negligibly small compared with $|\epsilon_1^{SC}(0)|$ and $\sigma_2^{res}(\omega)$ is a regular function at $\omega =$ 0 the equation above can be rewritten as $\Omega_{sp}^2 = \lim_{\omega \rightarrow 0}[-\omega^2 \epsilon_1^{SC}(\omega)]$. It is worth to be noted that the superfluid plasma frequency ($\Omega_{sp}$) can be independently obtained from a missing spectral weight of the full $\sigma^{SC}_1(\omega)$, which is the superfluid spectral weight located at $\omega =$ 0 and does not appear in a measured finite frequency region. The missing spectral weight can be obtained using the so-called Ferrell-Glover-Tinkham sum rule\cite{glover:1956,ferrell:1958,hwang:2007a}.

\section{Resulting optical self-energy data of model calculations}

\begin{figure}[!htbp]
  \vspace*{-0.3 cm}%
  \centerline{\includegraphics[width=6 in]{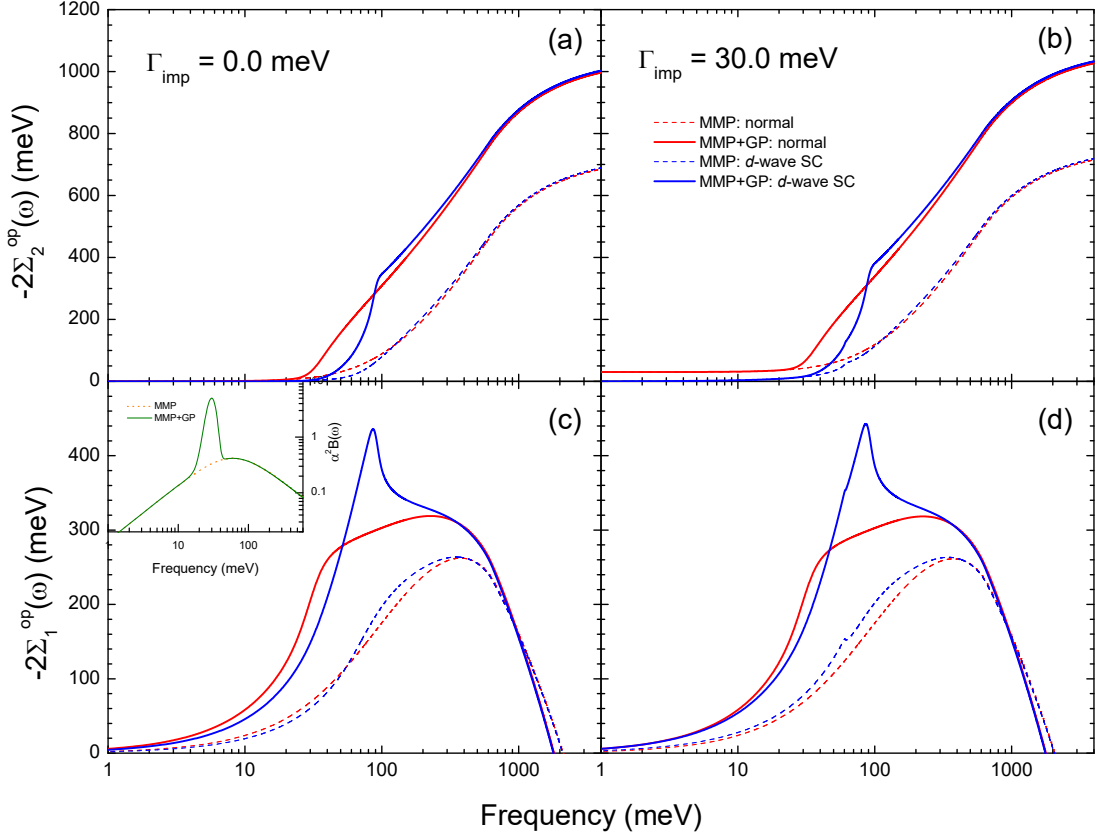}}%
  \vspace*{-0.3 cm}%
\caption{(Color online) (a) and (b) the calculated imaginary parts of the optical self-energies of the two (normal and $d$-wave SC) phases for two different impurity scattering rate cases: $\Gamma_{imp} =$ 0.0 and 30.0 meV, respectively. (c) and (d) the corresponding real parts of the optical self-energies of the two (normal and $d$-wave SC) phases for two different impurity scattering rate cases: $\Gamma_{imp} =$ 0.0 and 30.0 meV, respectively. In the inset of frame (c) two typical input model electron-boson spectral density functions, $\alpha^2B(\omega)$ (the MMP and MMP+GP models) are shown.}
 \label{fig1}
\end{figure}

For the theoretical calculations we consider two [normal and $d$-wave superconducting (SC)] phases with two typical input model electron-boson spectral density functions, $\alpha^2B(\omega)$\cite{hwang:2006,carbotte:2011}. One typical model $\alpha^2B(\omega)$ is for high temperature normal or highly overdoped SC phases\cite{hwang:2004,carbotte:2011} and is denoted as an MMP model, which was proposed by Millis, Monien, and Pines (MMP)\cite{millis:1990} and can be used for describing the antiferromagnetic spin-fluctuations. The other model $\alpha^2B(\omega)$ is for the low temperature phase below the coherence onset temperature\cite{basov:2005,carbotte:2011} and is denoted as an MMP+GP model, which consists of two components: the MMP model\cite{millis:1990} and a sharp Gaussian peak (GP), which can be used for describing the magnetic resonance mode observed by inelastic neutron scattering experiments\cite{rossat:1991,dai:1999,carbotte:1999,fong:2000,hwang:2004,stock:2004,hwang:2006}. We note that the temperature- and doping-dependent evolutions of $\alpha^2B(\omega)$ of hole-doped cuprates have been well-established by various experimental techniques including optical spectroscopy\cite{hwang:2004,hwang:2006,zasadzinski:2006,hwang:2007,dai:1999,fong:2000,carbotte:2011,johnson:2001,valla:2007,zhang:2008}. In the inset of Fig. \ref{fig1}(c) we show the two typical (MMP and MMP+GP) model $\alpha^2B(\omega)$. It is worth noting that the both model $\alpha^2B(\omega)$ are extended only up to 600 meV and above this energy they become zero. The MMP+GP model is shown in the solid olive line and can be written in a mathematical form as $\alpha^2B(\omega) = A_{SF} \: \omega/(\omega^2+\Omega_{SF}^2) +  A_{MR}\exp\{-(\omega-\Omega_0)^2/[2(W/2.35)^2]\}/[\sqrt{2\pi(W/2.35)^2}]$ where $A_{SF}$ (= 50 meV) and $\Omega_{SF}$ (= 60 meV) are the amplitude and the characteristic frequency of the MMP model, respectively, and $A_{MR}$ (= 50 meV), $W$ (= 10 meV), and $\Omega_0$ (= 30 meV) are the amplitude, width, and center frequency of the sharp GP, respectively. The MMP model is shown in the dotted orange line and can be also written in a mathematical form as $\alpha^2B(\omega) = A_{SF} \: \omega/(\omega^2+\Omega_{SF}^2)$ with the same amplitude and characteristic frequency of the MMP mode described above. The correlation constant ($\lambda$) of $\alpha^2B(\Omega)$ is an important and robust quantity, which can be a measure of the correlation strength\cite{hwang:2016a}, and can be calculated using $\lambda \equiv 2\int_0^{\omega_c} d\Omega\: \alpha^2B(\Omega)/\Omega$, where $\omega_c$ is a cutoff frequency, 600 meV for our cases. The estimated correlation constants of the MMP and MMP+GP models are 2.42 and 5.83, respectively.

In Fig. \ref{fig1}(a) and Fig. \ref{fig1}(b) we show calculated imaginary parts of the optical self-energy in a wide spectral range for two different input $\alpha^2B(\omega)$ (MMP and MMP+GP models) in two (normal and $d$-wave SC) phases with two different impurity scattering rates, $\Gamma_{imp} =$ 0.0 meV [frame (a)] and 30.0 meV [frame (b)]. Here the maximum SC gap ($\Delta_0$) is 30 meV. Since the input $\alpha^2B(\omega)$ is extended only up to 600 meV, above this energy, all the imaginary parts of the self-energies become rapidly saturated. For the $d$-wave SC phase with $\Gamma_{imp} =$ 30 meV, as one can expect from Eq. (3) significant impurity effects appear in the low frequency region below twice the maximum superconducting gap ($2\Delta_0 =$ 60 meV). For the normal phase the imaginary part of the optical self-energy is just vertically shifted by 30 meV in the whole spectral range compared with the corresponding self-energy with $\Gamma_{imp} =$ 0.0 meV since the impurity scattering is a (frequency-independent) constant, 30 meV. It should be noted that, in principle, the imaginary part of the optical self-energy should be saturated to $2\pi \times$the area under $\alpha^2B(\omega)$\cite{hwang:2008a}. For the $\Gamma_{imp} =$ 0.0 meV case the saturated values of the MMP and MMP+GP models are 724 meV and 1038 meV, respectively. Comparing the imaginary part of the optical self-energy of the MMP+GP model with that of the MMP model we can see that the sharp GP appears as a step-like feature\cite{hwang:2004} since the GP is very sharp. Since the height of the step-like feature will be around $2\pi \times$the area of the GP, from this height, the intensity of the sharp GP mode can be estimated\cite{hwang:2004}. The step-like features for two different (normal and SC) phases appear at different energies; in general, the feature appears near $\Omega_0$ for the normal phase\cite{carbotte:2005} and near $2\Delta_0+\Omega_0$ for the SC one\cite{carbotte:1990}.

In Fig. \ref{fig1}(c) and Fig. \ref{fig1}(d) we show corresponding real parts of the optical self-energy for $\Gamma_{imp} =$ 0.0 meV and 30.0 meV cases, respectively. These real parts of the optical self-energy are obtained from the calculated imaginary parts in a wide spectral range from 0 to 7000 meV using the Kramers-Kronig relation, Eq. (4). In principle, the real part should carry additional complementary information. This real part of the optical self-energy is closely related to the mass enhancement function of charge carriers as $-2\Sigma^{op}_1(\omega) = [m_{op}^*(\omega)/m_b -1]\omega$. For the broad MMP model, two self-energies in normal and $d$-wave SC phases are quite similar to each other, which means that these two different material phases are not very well-resolved. In contrast, for the MMP+GP model, due to the sharp GP the (normal and SC) material phases are well-resolved in low frequency region (near $\Omega_0 + 2\Delta_0$). Therefore, the doping- and temperature-dependent evolutions of the GP can be clearly investigated from measured optical spectra\cite{hwang:2004,hwang:2006,hwang:2007}.

\section{Revisit the proposed method with a different approach}

Dordevic {\it et al.}\cite{dordevic:2014} proposed a method to obtain the optical self-energy of quasiparticles (or residual electrons) in the superconducting state. We investigated the method in the correlated electron formalism introduced in the previous sections. We obtained the optical self-energies of the residual electrons for the typical MMP and MMP+PG models in the $d$-wave SC phase using the reverse process, compared and discussed the resulting residual (or partial) self-energies with the corresponding full SC self-energies, and provided some important comments on the proposed method.

\begin{figure}[!htbp]
  \vspace*{-0.3 cm}%
  \centerline{\includegraphics[width=5.3 in]{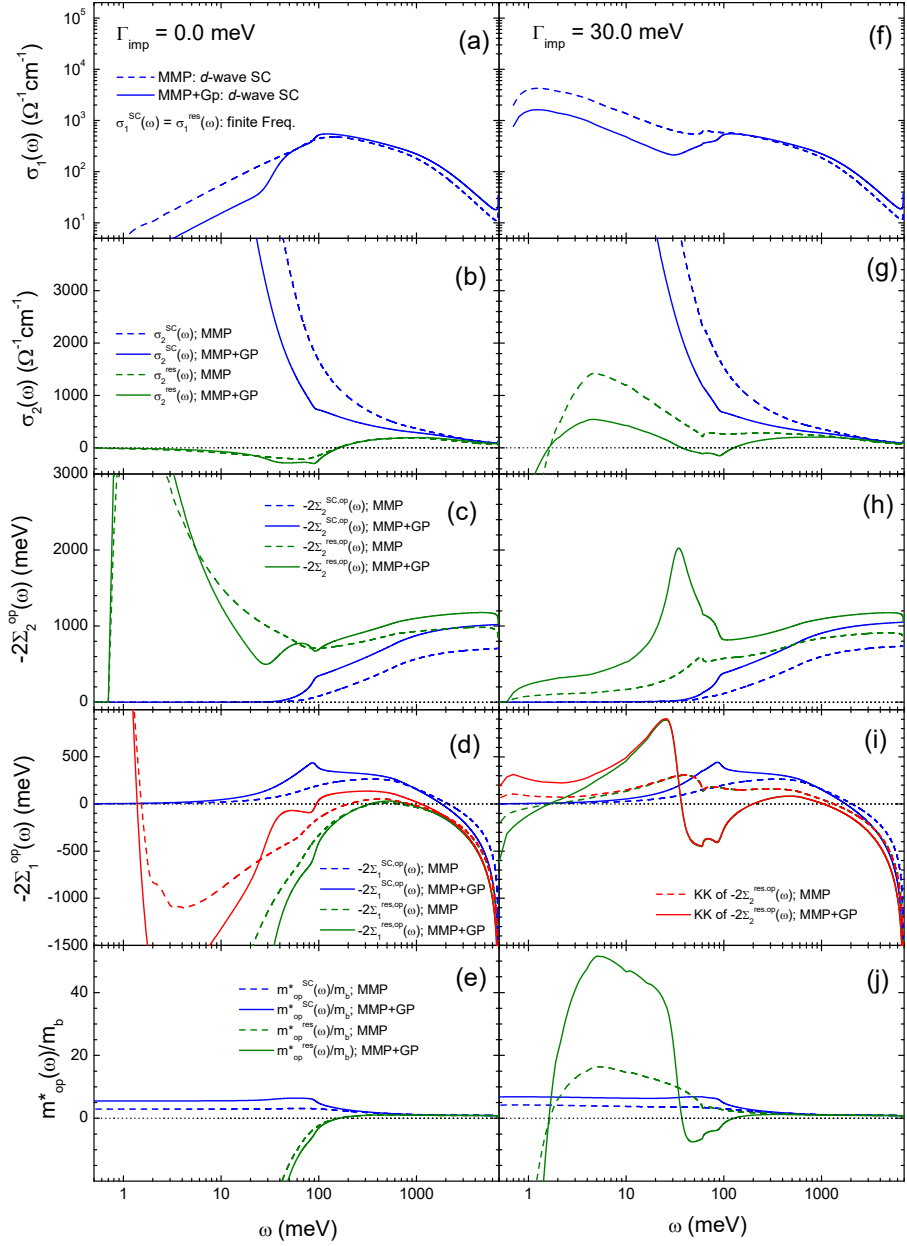}}%
  \vspace*{-0.7 cm}%
\caption{(Color online) In left column from the top frame (a) to the bottom one (e) the real part of the optical conductivity, the imaginary part of the optical conductivity, the imaginary part of optical self-energy, the real part of the optical self-energy, and the mass enhancement function of the MMP and MMP+GP models in $d$-wave SC phase for $\Gamma_{imp}=$ 0.0 meV. In the right column from the top frame (f) to the bottom one (j) the same quantities in the same order for $\Gamma_{imp}=$ 30.0 meV. We show both residual and full SC quantities for comparison. In (d) and (i) we also show the KK transformed residual optical self-energies [KK of $-2\Sigma_2^{res,op}(\omega)$] obtained from the imaginary parts of the residual self-energies using the Kramers-Kronig relation [Eq. (4)] for $\Gamma_{imp}=$ 0.0 meV and $\Gamma_{imp}=$ 30.0 meV, respectively.}
 \label{fig2}
\end{figure}

Here we describe recent Dordevic {\it et al.}'s work\cite{dordevic:2014}. The authors raised a concern on the usual extended Drude model analysis of optical data in the $d$-wave superconducting phase and suggested that the optical self-energy of quasiparticles (or the residual unpaired electrons) in the $d$-wave SC phase can be obtained using the residual (or partial) optical conductivity [$\tilde{\sigma}^{res}(\omega)\equiv\sigma_1^{res}(\omega)+i\sigma_2^{res}(\omega)$], which was defined in Eq. (6). The authors also proposed that to determine the intrinsic quasiparticle (or residual electron) properties in the SC phase one must use the residual optical conductivity instead of the full SC optical conductivity [$\tilde{\sigma}^{SC}(\omega)\equiv\sigma_1^{SC}(\omega)+i\sigma_2^{SC}(\omega)$] in the extended Drude formalism. In this case since only the residual electrons are considered the plasma frequency in the extended Drude model should be modified as $\Omega_p^{res} = \sqrt{\Omega_p^2 - \Omega_{sp}^2}$, where $\Omega_p^{res}$ is the modified (or residual) plasma frequency. Then the residual (or partial) optical self-energy, $-2\tilde{\Sigma}^{res,op}(\omega)$ ($\equiv \:-2\Sigma_1^{res,op}(\omega)+ i[-2\Sigma_2^{res,op}(\omega)]$), can be written as
\begin{equation}\label{}
-2\tilde{\Sigma}^{res,op}(\omega) = i \frac{[\Omega_p^{res}]^2}{4 \pi}\frac{1}{\tilde{\sigma}^{res}(\omega)}-\omega.
\end{equation}
Here the residual (or partial) optical self-energy is described in terms of the residual (or partial) optical conductivity. In this section we will demonstrate that the real and imaginary parts of the residual optical self-energy [$-2\tilde{\Sigma}^{res,op}(\omega)$] do not form a KK pair. We should note that the real and imaginary parts of the full SC self-energy [$-2\tilde{\Sigma}^{SC,op}(\omega)$], which is described in terms of the full optical SC conductivity [$\tilde{\sigma}^{SC}(\omega)$], are known to form a KK pair\cite{hwang:2004}. This is one of the most important differences between the two optical self-energies: the partial one [$-2\tilde{\Sigma}^{res,op}(\omega)$] and the full one [$-2\tilde{\Sigma}^{SC,op}(\omega)$]. Therefore, the residual (or partial) optical self-energy may not be a reliable (or well-defined) optical quantity.

To investigate the two optical self-energies more in detail, first we calculated the full SC optical conductivities of the MMP and MMP+GP models in the $d$-wave SC phase for both $\Gamma_{imp}=$ 0.0 and 30.0 meV cases from the corresponding full SC optical self-energies shown in Fig. \ref{fig1} through the extended Drude formalism [Eq. (5)]. Here we used the plasma frequency ($\Omega_p$) of 2000 meV. Then using the Kramers-Kronig relation we obtained the imaginary part of the residual optical conductivity [$\sigma_2^{res}(\omega)$] from the real part of the residual optical conductivity [$\sigma_1^{res}(\omega)$]. To get the residual plasma frequency ($\Omega_p^{res}$) we obtained the superfluid plasma frequencies ($\Omega_{sp}$) for all four cases using Eq. (7). The obtained $\Omega_{sp}$ of the MMP model were 1158 meV and 965 meV for $1/\tau_{imp} =$ 0 meV and 30 meV, respectively and those of the MMP+GP model were 843 meV and 760 meV for $1/\tau_{imp} =$ 0.0 meV and 30.0 meV, respectively. We note that the impurity scattering rate reduces the superfluid plasma frequency\cite{hwang:2015a}. Finally we obtained the residual (or partial) optical self-energy [$-2\tilde{\Sigma}^{res,op}(\omega)$] from the residual optical conductivity [$\tilde{\sigma}^{res}(\omega)$] using the extended Drude model, Eq. (9) with appropriate residual plasma frequencies for all four cases, i.e., $\Omega_p^{res} = \sqrt{\Omega_p^2-\Omega_{sp}^2}$. The used residual plasma frequencies of the MMP model were 1630 meV and 1752 meV for $1/\tau_{imp} =$ 0.0 meV and 30.0 meV, respectively, and those of MMP+GP model were 1814 meV and 1850 meV for $1/\tau_{imp} =$ 0.0 meV and 30.0 meV, respectively.

Now we compared the resulting residual (or partial) self-energy [$-2\tilde{\Sigma}^{res,op}(\omega)$] with the full SC self-energy [$-2\tilde{\Sigma}^{SC,op}(\omega)$], which are taken from Fig. \ref{fig1}. We show all results of our calculations in Fig. \ref{fig2}.

In Fig. \ref{fig2}(a) and \ref{fig2}(f) we show $\sigma_1^{SC}(\omega)$ of the MMP (the dashed lines) and MMP+GP (the solid lines) models in the $d$-wave SC phase, respectively, for $\Gamma_{imp}=$ 0.0 meV and 30.0 meV cases. As we already have mentioned $\sigma_1^{SC}(\omega) = \sigma_1^{res}(\omega)$ for finite frequency, $\omega > 0$ but $\sigma_1^{SC}(0) \neq \sigma_1^{res}(0)$. We note that the optical conductivity consists of two (coherent and incoherent) components. While almost no coherent components appear in the residual conductivities of the MMP and MMP+GP models for $1/\tau_{imp} = 0.0$ meV case most of the coherent components appear in the residual conductivities for $1/\tau_{imp} = 30.0$ meV case\cite{hwang:2015a}. It should be noted that the residual electrons in the MMP and MMP+GP models seem to be governed by the $d$-wave SC gap for both $1/\tau_{imp} =$ 0.0 meV and 30.0 meV cases; there are non-zero spectral weights below two times of the SC gap ($2\Delta_0$ = 60 meV).

In Fig. \ref{fig2}(b) and \ref{fig2}(g) we show corresponding imaginary parts of the residual and full SC optical conductivities [$\sigma_2^{res}(\omega)$ and $\sigma_2^{SC}(\omega)$] of the MMP (the dashed lines) and MMP+GP (the solid lines) models in the $d$-wave SC phase, respectively, for $\Gamma_{imp}=$ 0.0 meV and 30.0 meV cases. Here $\sigma_2^{res}(\omega)$ are obtained from $\sigma_1^{res}(\omega)$ using the Kramers-Kronig relation. The obtained residual $\sigma_2^{res}(\omega)$ (the olive lines) are completely different from the corresponding $\sigma_2^{SC}(\omega)$ (the blue lines) and become negative in some spectral regions, which is closely related to negative optical effective mass [see Fig. \ref{fig2}(e) and \ref{fig2}(j)]. We note that in low frequency region $\sigma_2^{SC}(\omega)$ clearly shows the response from electrons involved in the superconductivity.

In Fig. \ref{fig2}(c) and \ref{fig2}(h) we show corresponding imaginary parts of the two (residual and full) optical self-energies [$-2\Sigma_2^{res,op}(\omega)$ and $-2\Sigma_2^{SC,op}(\omega)$], respectively, for $\Gamma_{imp}=$ 0.0 meV and 30.0 meV cases. In Fig. \ref{fig2}(c) both $-2\Sigma_2^{res,op}(\omega)$ of the MMP and MMP+PG models with $\Gamma_{imp} =$ 0.0 meV show very strong peaks well below two times of the SC gap (2$\Delta_0 =$ 60 meV). In Fig. \ref{fig2}(h) $-2\Sigma_2^{res,op}(\omega)$ of the MMP+GP model with $\Gamma_{imp} =$ 30.0 meV also show a quite strong peak below two times of the SC gap. These strong peaks seem to be unphysical since they cannot be simulated using the generalized Allen's formula in the normal phase (i.e, with a constant density of states)\cite{hwang:2013a}; to simulate such strong peaks, one must introduce very strong intensity modulations in the density of states\cite{hwang:2013a}.

In Fig. \ref{fig2}(d) and \ref{fig2}(i) we show corresponding real parts of the optical self-energies [$-2\Sigma_1^{res,op}(\omega)$ and $-2\Sigma_1^{SC,op}(\omega)$], respectively, for $\Gamma_{imp}=$ 0.0 meV and 30.0 meV cases. All four residual (or partial) self-energies [$-2\Sigma_1^{res,op}(\omega)$] are significantly different from the corresponding full SC self-energies [$-2\Sigma_1^{SC,op}(\omega)$]. We also show a new set of real parts of the optical self-energies in the solid (MMP+GP model) and dashed (MMP model) red lines, which are obtained from $-2\Sigma_2^{res,op}(\omega)$ using the Kramers-Kronig relation, Eq. (4). We denote them as KK of $-2\Sigma_2^{res,op}(\omega)$. Comparing the KK of $-2\Sigma_2^{res,op}(\omega)$ with $-2\Sigma_1^{res,op}(\omega)$ we can clearly see discrepancies between them, which mean that $-2\Sigma_1^{res,op}(\omega)$ and $-2\Sigma_2^{res,op}(\omega)$ do not form a KK pair. Interestingly, the degree of the discrepancy depends upon the intensity of the impurity scattering rate. This discrepancy may indicate that the residual (or partial) optical self-energy cannot be a reliable optical quantity. We observed that all the real parts of the residual optical self-energies [$-2\Sigma_1^{res,op}(\omega)$] of the MMP and MMP+GP models become negative in low frequency regions; this seems to be unphysical since negative values of the real part of the optical self-energy cannot be physically interpreted.

Finally, in Fig. \ref{fig2}(e) and \ref{fig2}(j) we show the calculated residual and full SC mass enhancement functions [$m*_{op}^{res}(\omega)/m_b$ and $m*_{op}^{SC}(\omega)/m_b$], respectively, for $\Gamma_{imp}=$ 0.0 meV and 30.0 meV cases. All four $m*_{op}^{res}(\omega)/m_b$ of the MMP (the dashed olive lines) and MMP+GP models (the solid olive lines) seem to be unphysical since they become strongly negative in low frequency regions: for $\Gamma_{imp}=$ 0.0 meV case they become negative below $\sim$157 meV and for $\Gamma_{imp}=$ 30.0 meV case they become negative below $\sim$1.5 meV.

From our theoretical investigation, we found that the proposed method by Dordevic {\it et al.} contains two serious problems. One problem is associated with peaks in the imaginary part of the residual (or partial) optical self-energy; this method cannot be applied since the peaks cannot be simulated with a constant density of states (i.e., the normal phase)\cite{hwang:2013a}. The other problem is that the real part of the residual optical self-energy and the corresponding effective mass function exhibit unphysical negative values in the low frequency region.

There has been a previously published literature\cite{schachinger:2017} motivated by the attempt by Dordevic {\it et al.}\cite{dordevic:2014}. In their paper Schachinger and Carbotte calculated four optical self-energies (the optical self-energy of Bogoliubov quasiparticles (BQPs), the residual optical self-energy, and the optical self-energies for both SC and normal states) of Pb using a formal approach, which is different from ours. The authors compared their four optical self-energies and found that the residual self-energy does not reflect purely the BQP self-energy since the imaginary part of the residual self-energy is strongly deviated from those of other three self-energies in high frequency region well above the SC gap (or 8 times of the SC gap) while the imaginary parts of other three self-energies merge in the high frequency. Furthermore, they focused on optical properties of BQPs. However, the authors did not mention the unphysical negative effective mass function which we described in our paper.

\section{Application of the method to measured optical data and discussion}

We applied the proposed method\cite{dordevic:2014} described in the previous section to real material systems, i.e., optimally doped and overdoped Bi$_2$Sr$_2$CaCu$_2$O$_{8+\delta}$ (Bi2212) systems. The optimally doped Bi2212 with the superconducting transition temperature, $T_c =$ 96 K is denoted as Bi2212-OPT96 and the overdoped Bi2212 with $T_c =$ 60 K as Bi2212-OD60. It is worth noting that since in these Bi2212 cuprate systems the extended Drude mode extends up to very high energy, $\sim$2.0 eV\cite{hwang:2007a} their optical self-energies are reliable in a wide spectral range up to near 2.0 eV.

\begin{figure}[!htbp]
  \vspace*{-0.3 cm}%
  \centerline{\includegraphics[width=6 in]{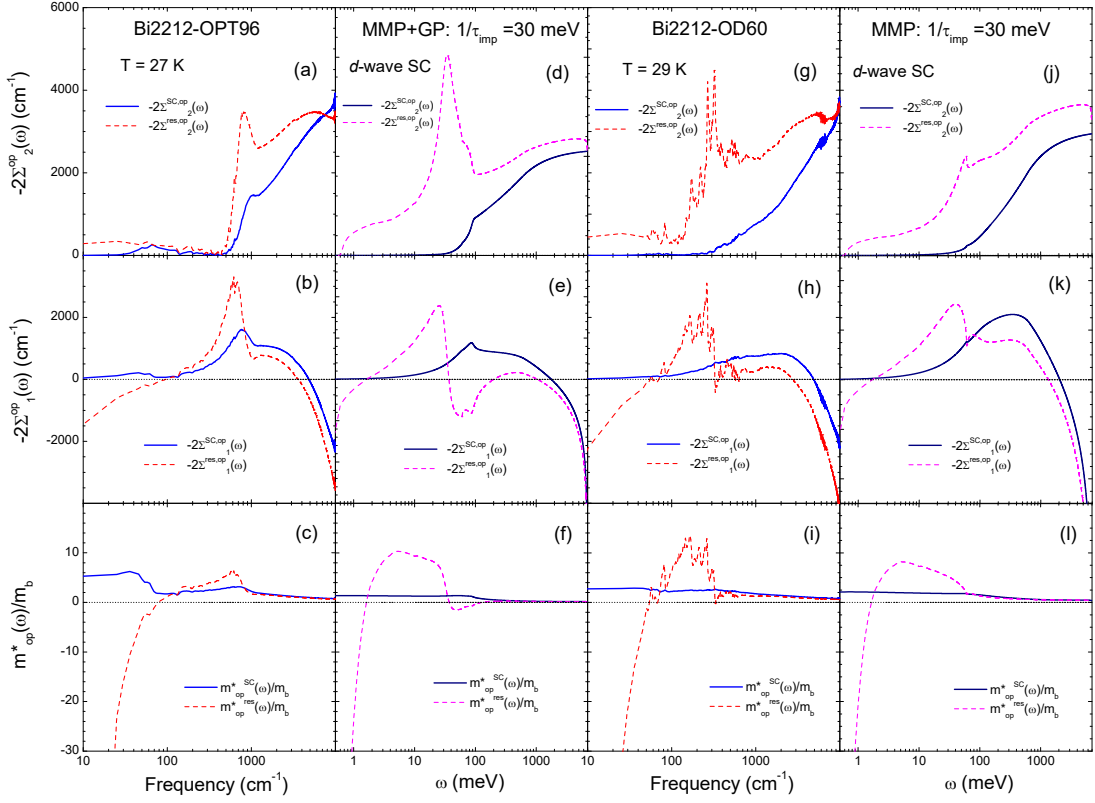}}%
  \vspace*{-0.5 cm}%
\caption{(Color online) Results of application of the new method to two cuprate systems: Bi2212-OPT96 and Bi2212-OD60. (a)-(c) The calculated residual and full SC optical self-energies [$-2\tilde{\Sigma}^{res,op}(\omega)$ and $-2\tilde{\Sigma}^{SC,op}(\omega)$] and the mass enhancement functions [$m*_{op}^{res}(\omega)/m_b$ and $m*_{op}^{SC}(\omega)/m_b$] of Bi2212-OPT96 at $T =$ 27 K. (d)-(f) The obtained residual and full SC optical self-energies and the mass enhancement functions of the MMP+GP model. (g)-(i) The resulting residual and full SC optical self-energies and the mass enhancement functions of Bi2212-OD60 at $T =$ 29 K. (j)-(l) the resulting residual and full SC optical self-energies and the mass enhancement function of the MMP model.}
 \label{fig3}
\end{figure}

In Fig. \ref{fig3} (a)-(c) we show the resulting residual (the dashed red lines) and full SC (the solid blue lines) self-energies [$-2\tilde{\Sigma}^{res,op}(\omega)$ and $-2\tilde{\Sigma}^{SC,op}(\omega)$] and the residual (the dashed red line) and full SC (the solid bule line) mass enhancement functions [$m*_{op}^{res}(\omega)/m_b$ and $m*_{op}^{SC}(\omega)/m_b$] of Bi2212-OPT96 at $T =$ 27 K. In Fig. \ref{fig3} (d)-(f) we show the corresponding quantities of the MMP+GP model in the $d$-wave SC phase with $1/\tau_{imp} =$ 30.0 meV. We compared these two sets of the measured optical self-energies [$-2\tilde{\Sigma}^{res,op}(\omega)$ and $-2\tilde{\Sigma}^{SC,op}(\omega)$] of Bi2212-OPT96 at $T =$ 27 K with the corresponding theoretically calculated optical self-energies of the MMP+GP model in the $d$-wave SC phase since both (Bi2212-OPT96 at 27 K and the MMP+GP model in the $d$-wave SC phase) were known to belong to the same region in the generic $T-p$ phase diagram of hole-doped cuprates\cite{hwang:2018}. The corresponding residual and full SC optical quantities of the two (real Bi2212-OPT96 and theoretical MMP+GP model) systems agree qualitatively well each other. We note that the full SC optical quantities of real Bi2212-OPT96 and theoretical MMP+GP model systems agree better than the corresponding residual quantities. The imaginary part of the theoretical residual self-energy shows a stronger peak around 34 meV than that of the experimental residual self-energy. Due to this stronger peak the corresponding real part of the theoretical self-energy shows negative values in a region between 37 meV and 200 meV while the real part of the experimental self-energy does not show negative values near its peak frequency. The theoretical residual mass enhancement function shows a similar negative region as shown in Fig. \ref{fig3}(f). The (experimental and theoretical) real parts of the residual self-energies and residual mass enhancement functions become negative in low-energy regions, which is unphysical since the negative values cannot be physically interpreted.

In Fig. \ref{fig3} (g)-(i) we show the resulting residual (the dashed red lines) and full SC (the solid blue lines) optical self-energies and the residual (the dashed red lines) and full (the solid blue lines) SC mass enhancement functions of Bi2212-OD60 at $T =$ 29 K. In Fig. \ref{fig3} (j)-(l) we display the corresponding quantities of the MMP model in the $d$-wave SC phase with $1/\tau_{imp} =$ 30.0 meV. We compared the residual and full SC optical self-energies of Bi2212-OD60 at $T =$ 29 K with those of the MMP model in the $d$-wave SC phase since both (Bi2212-OD60 at 29 K and the MMP model in the $d$-wave SC phase) belong to the same region in the generic $T-p$ phase diagram of hole-doped cuprates\cite{hwang:2018}. As we can see in the figure, the corresponding optical quantities of the measured Bi2212-OD60 and calculated MMP model systems are qualitatively very similar to each other. The real parts of the two (measured and calculated) residual self-energies become negative in low-energy region. The corresponding residual mass enhancement functions become also negative in a similar spectral region. As we mentioned previously, the negative values in the real part of the residual self-energies and residual mass enhancement functions cannot be physically interpreted.

Our comparison indicates that when we applied the proposed method to experimentally measured spectra, the resulting residual (or partial) optical self-energy shows the same problematic (or unphysical) features evident in the theoretically calculated residual optical self-energy. The problematic features were the peaks in the imaginary part of the residual self-energy, and the negative values of the real part of the residual optical self-energy in the low frequency region. In contrast, the full SC self-energy defined by the extended Drude model does not exhibit these problematic features; this full SC optical self-energy can be a reliable optical quantity.

\section{Conclusions}

Using a reverse process\cite{hwang:2015a} we investigated a new proposed method\cite{dordevic:2014}, that was suggested to study the residual unpaired electrons of cuprates in the superconducting phase. First, we obtained the optical self-energy in two different material phases (normal and $d$-wave SC) with the two typical MMP and MMP+GP model electron-boson spectral density functions, $\alpha^2B(\omega)$, using the generalized Allen's formulas [Eq. (1)] and the Kramers–Kronig relation [Eq. (4)]. Subsequently, we obtained the optical conductivity from the obtained optical self-energy using the extended Drude formalism. We calculated the so-called residual (or partial) optical self-energy [$-2\tilde{\Sigma}^{res,op}(\omega)$] of the MMP and MMP+GP models in the $d$-wave SC phase using the new proposed method by Dordevic {\it et al.}\cite{dordevic:2014}. The obtained imaginary part of the residual optical self-energies demonstrated unphysical peaks. We found that the real and imaginary parts of the residual self-energy did not form a KK pair. Therefore, this residual self-energy was not an appropriate optical quantity. Additionally, the real part of the residual self-energy [$-2\Sigma_1^{res,op}(\omega)$] and the corresponding residual mass enhancement function [$m*_{op}^{res}(\omega)/m_b$] became negative in the low-energy region, which were unphysical features. We also applied the method to two Bi2212 cuprate systems (Bi2212-OPT96 and Bi2212-OD60) in the SC phase. We observed that the obtained residual self-energies and residual mass enhancement functions of Bi2212-OP96 at 27 K and Bi2212-OD60 at 29 K demonstrated similar unphysical features (peaks in $-2\Sigma_2^{res,op}(\omega)$ and negative values of $-2\Sigma_1^{res,op}(\omega)$ and $m*_{op}^{res}(\omega)/m_b$ in the low frequency region) as in the corresponding quantities of the MMP+GP and MMP models in the $d$-wave SC phase. From our results, we concluded that the proposed method could be used for analyzing the measured optical data of Bi2212 cuprate systems in the SC phase; the residual (or partial) optical self-energy was not a reliable optical quantity because it contained unphysical features. Instead, the full SC optical self-energy was a reliable optical quantity that did not contain unphysical features. In our opinion, the full optical self-energy could not be decomposed even though the full optical conductivity could be decomposed because they were connected non-linearly by the extended Drude model. We hope that our study clarifies an unclear issue related to the extended Drude model analysis of the superconducting optical data of strongly correlated electron systems including cuprates and iron pnictides.

%
\acknowledgments This paper was supported by the National Research Foundation of Korea (NRFK Grant No. 2017R1A2B4007387).

\bibliographystyle{apsrev4-1}
\bibliography{bib}

\begin{thebibliography}{46}%
\makeatletter
\providecommand \@ifxundefined [1]{%
 \@ifx{#1\undefined}
}%
\providecommand \@ifnum [1]{%
 \ifnum #1\expandafter \@firstoftwo
 \else \expandafter \@secondoftwo
 \fi
}%
\providecommand \@ifx [1]{%
 \ifx #1\expandafter \@firstoftwo
 \else \expandafter \@secondoftwo
 \fi
}%
\providecommand \natexlab [1]{#1}%
\providecommand \enquote  [1]{``#1''}%
\providecommand \bibnamefont  [1]{#1}%
\providecommand \bibfnamefont [1]{#1}%
\providecommand \citenamefont [1]{#1}%
\providecommand \href@noop [0]{\@secondoftwo}%
\providecommand \href [0]{\begingroup \@sanitize@url \@href}%
\providecommand \@href[1]{\@@startlink{#1}\@@href}%
\providecommand \@@href[1]{\endgroup#1\@@endlink}%
\providecommand \@sanitize@url [0]{\catcode `\\12\catcode `\$12\catcode
  `\&12\catcode `\#12\catcode `\^12\catcode `\_12\catcode `\%12\relax}%
\providecommand \@@startlink[1]{}%
\providecommand \@@endlink[0]{}%
\providecommand \url  [0]{\begingroup\@sanitize@url \@url }%
\providecommand \@url [1]{\endgroup\@href {#1}{\urlprefix }}%
\providecommand \urlprefix  [0]{URL }%
\providecommand \Eprint [0]{\href }%
\providecommand \doibase [0]{http://dx.doi.org/}%
\providecommand \selectlanguage [0]{\@gobble}%
\providecommand \bibinfo  [0]{\@secondoftwo}%
\providecommand \bibfield  [0]{\@secondoftwo}%
\providecommand \translation [1]{[#1]}%
\providecommand \BibitemOpen [0]{}%
\providecommand \bibitemStop [0]{}%
\providecommand \bibitemNoStop [0]{.\EOS\space}%
\providecommand \EOS [0]{\spacefactor3000\relax}%
\providecommand \BibitemShut  [1]{\csname bibitem#1\endcsname}%
\let\auto@bib@innerbib\@empty
\bibitem [{\citenamefont {G\"{o}tze}\ and\ \citenamefont
  {W\"{o}lfle}(1972)}]{gotze:1972}%
  \BibitemOpen
  \bibfield  {author} {\bibinfo {author} {\bibfnamefont {W.}~\bibnamefont
  {G\"{o}tze}}\ and\ \bibinfo {author} {\bibfnamefont {P.}~\bibnamefont
  {W\"{o}lfle}},\ }\href@noop {} {\bibfield  {journal} {\bibinfo  {journal}
  {Phys. Rev. B}\ }\textbf {\bibinfo {volume} {6}},\ \bibinfo {pages} {1226}
  (\bibinfo {year} {1972})}\BibitemShut {NoStop}%
\bibitem [{\citenamefont {Hwang}\ \emph {et~al.}(2004)\citenamefont {Hwang},
  \citenamefont {Timusk},\ and\ \citenamefont {Gu}}]{hwang:2004}%
  \BibitemOpen
  \bibfield  {author} {\bibinfo {author} {\bibfnamefont {J.}~\bibnamefont
  {Hwang}}, \bibinfo {author} {\bibfnamefont {T.}~\bibnamefont {Timusk}}, \
  and\ \bibinfo {author} {\bibfnamefont {G.~D.}\ \bibnamefont {Gu}},\
  }\href@noop {} {\bibfield  {journal} {\bibinfo  {journal} {Nature (London)}\
  }\textbf {\bibinfo {volume} {427}},\ \bibinfo {pages} {714} (\bibinfo {year}
  {2004})}\BibitemShut {NoStop}%
\bibitem [{\citenamefont {Carbotte}\ \emph {et~al.}(2005)\citenamefont
  {Carbotte}, \citenamefont {Schachinger},\ and\ \citenamefont
  {Hwang}}]{carbotte:2005}%
  \BibitemOpen
  \bibfield  {author} {\bibinfo {author} {\bibfnamefont {J.~P.}\ \bibnamefont
  {Carbotte}}, \bibinfo {author} {\bibfnamefont {E.}~\bibnamefont
  {Schachinger}}, \ and\ \bibinfo {author} {\bibfnamefont {J.}~\bibnamefont
  {Hwang}},\ }\href@noop {} {\bibfield  {journal} {\bibinfo  {journal} {Phys.
  Rev. B}\ }\textbf {\bibinfo {volume} {71}},\ \bibinfo {pages} {054506}
  (\bibinfo {year} {2005})}\BibitemShut {NoStop}%
\bibitem [{\citenamefont {Hwang}\ \emph
  {et~al.}(2007{\natexlab{a}})\citenamefont {Hwang}, \citenamefont {Nicol},
  \citenamefont {Timusk}, \citenamefont {Knigavko},\ and\ \citenamefont
  {Carbotte}}]{hwang:2007b}%
  \BibitemOpen
  \bibfield  {author} {\bibinfo {author} {\bibfnamefont {J.}~\bibnamefont
  {Hwang}}, \bibinfo {author} {\bibfnamefont {E.~J.}\ \bibnamefont {Nicol}},
  \bibinfo {author} {\bibfnamefont {T.}~\bibnamefont {Timusk}}, \bibinfo
  {author} {\bibfnamefont {A.}~\bibnamefont {Knigavko}}, \ and\ \bibinfo
  {author} {\bibfnamefont {J.~P.}\ \bibnamefont {Carbotte}},\ }\href@noop {}
  {\bibfield  {journal} {\bibinfo  {journal} {Phys. Rev. Lett.}\ }\textbf
  {\bibinfo {volume} {98}},\ \bibinfo {pages} {207002} (\bibinfo {year}
  {2007}{\natexlab{a}})}\BibitemShut {NoStop}%
\bibitem [{\citenamefont {Kotliar}\ and\ \citenamefont
  {Vollhardt}(2004)}]{kotlier:2004}%
  \BibitemOpen
  \bibfield  {author} {\bibinfo {author} {\bibfnamefont {G.}~\bibnamefont
  {Kotliar}}\ and\ \bibinfo {author} {\bibfnamefont {D.}~\bibnamefont
  {Vollhardt}},\ }\href@noop {} {\bibfield  {journal} {\bibinfo  {journal}
  {Phys. Today}\ }\textbf {\bibinfo {volume} {57}},\ \bibinfo {pages} {53}
  (\bibinfo {year} {2004})}\BibitemShut {NoStop}%
\bibitem [{\citenamefont {Webb}\ \emph {et~al.}(1986)\citenamefont {Webb},
  \citenamefont {Sievers},\ and\ \citenamefont {Mihalisin}}]{webb:1986}%
  \BibitemOpen
  \bibfield  {author} {\bibinfo {author} {\bibfnamefont {B.~C.}\ \bibnamefont
  {Webb}}, \bibinfo {author} {\bibfnamefont {A.~J.}\ \bibnamefont {Sievers}}, \
  and\ \bibinfo {author} {\bibfnamefont {T.}~\bibnamefont {Mihalisin}},\
  }\href@noop {} {\bibfield  {journal} {\bibinfo  {journal} {Phys. Rev. Lett.}\
  }\textbf {\bibinfo {volume} {57}},\ \bibinfo {pages} {1951} (\bibinfo {year}
  {1986})}\BibitemShut {NoStop}%
\bibitem [{\citenamefont {Sulewski}\ \emph {et~al.}(1988)\citenamefont
  {Sulewski}, \citenamefont {Sievers}, \citenamefont {Maple}, \citenamefont
  {Torikachvili}, \citenamefont {Smith},\ and\ \citenamefont
  {Fisk}}]{sulewski:1988}%
  \BibitemOpen
  \bibfield  {author} {\bibinfo {author} {\bibfnamefont {P.~E.}\ \bibnamefont
  {Sulewski}}, \bibinfo {author} {\bibfnamefont {A.~J.}\ \bibnamefont
  {Sievers}}, \bibinfo {author} {\bibfnamefont {M.~B.}\ \bibnamefont {Maple}},
  \bibinfo {author} {\bibfnamefont {M.~S.}\ \bibnamefont {Torikachvili}},
  \bibinfo {author} {\bibfnamefont {J.~L.}\ \bibnamefont {Smith}}, \ and\
  \bibinfo {author} {\bibfnamefont {Z.}~\bibnamefont {Fisk}},\ }\href@noop {}
  {\bibfield  {journal} {\bibinfo  {journal} {Phys. Rev. B}\ }\textbf {\bibinfo
  {volume} {38}},\ \bibinfo {pages} {5338} (\bibinfo {year}
  {1988})}\BibitemShut {NoStop}%
\bibitem [{\citenamefont {Thomas}\ \emph {et~al.}(1988)\citenamefont {Thomas},
  \citenamefont {Orenstein}, \citenamefont {Rapkine}, \citenamefont {Capizzi},
  \citenamefont {Millis}, \citenamefont {Bhatt}, \citenamefont {Schneemeyer},\
  and\ \citenamefont {Waszczak}}]{thomas:1988}%
  \BibitemOpen
  \bibfield  {author} {\bibinfo {author} {\bibfnamefont {G.~A.}\ \bibnamefont
  {Thomas}}, \bibinfo {author} {\bibfnamefont {J.}~\bibnamefont {Orenstein}},
  \bibinfo {author} {\bibfnamefont {D.~H.}\ \bibnamefont {Rapkine}}, \bibinfo
  {author} {\bibfnamefont {M.}~\bibnamefont {Capizzi}}, \bibinfo {author}
  {\bibfnamefont {A.~J.}\ \bibnamefont {Millis}}, \bibinfo {author}
  {\bibfnamefont {R.~N.}\ \bibnamefont {Bhatt}}, \bibinfo {author}
  {\bibfnamefont {L.~F.}\ \bibnamefont {Schneemeyer}}, \ and\ \bibinfo {author}
  {\bibfnamefont {J.~V.}\ \bibnamefont {Waszczak}},\ }\href@noop {} {\bibfield
  {journal} {\bibinfo  {journal} {Phys. Rev. Lett.}\ }\textbf {\bibinfo
  {volume} {61}},\ \bibinfo {pages} {1313} (\bibinfo {year}
  {1988})}\BibitemShut {NoStop}%
\bibitem [{\citenamefont {Awasthi}\ \emph {et~al.}(1993)\citenamefont
  {Awasthi}, \citenamefont {Degiorgi}, \citenamefont {Gruner}, \citenamefont
  {Dalichaouch},\ and\ \citenamefont {B.Maple}}]{awasthi:1993}%
  \BibitemOpen
  \bibfield  {author} {\bibinfo {author} {\bibfnamefont {A.~M.}\ \bibnamefont
  {Awasthi}}, \bibinfo {author} {\bibfnamefont {L.}~\bibnamefont {Degiorgi}},
  \bibinfo {author} {\bibfnamefont {G.}~\bibnamefont {Gruner}}, \bibinfo
  {author} {\bibfnamefont {Y.}~\bibnamefont {Dalichaouch}}, \ and\ \bibinfo
  {author} {\bibfnamefont {M.}~\bibnamefont {B.Maple}},\ }\href@noop {}
  {\bibfield  {journal} {\bibinfo  {journal} {Phys. Rev. B}\ }\textbf {\bibinfo
  {volume} {48}},\ \bibinfo {pages} {10692} (\bibinfo {year}
  {1993})}\BibitemShut {NoStop}%
\bibitem [{\citenamefont {Puchkov}\ \emph {et~al.}(1996)\citenamefont
  {Puchkov}, \citenamefont {Basov},\ and\ \citenamefont
  {Timusk}}]{puchkov:1996}%
  \BibitemOpen
  \bibfield  {author} {\bibinfo {author} {\bibfnamefont {A.~V.}\ \bibnamefont
  {Puchkov}}, \bibinfo {author} {\bibfnamefont {D.~N.}\ \bibnamefont {Basov}},
  \ and\ \bibinfo {author} {\bibfnamefont {T.}~\bibnamefont {Timusk}},\
  }\href@noop {} {\bibfield  {journal} {\bibinfo  {journal} {J. Phys.: Cond.
  Matter}\ }\textbf {\bibinfo {volume} {8}},\ \bibinfo {pages} {10049}
  (\bibinfo {year} {1996})}\BibitemShut {NoStop}%
\bibitem [{\citenamefont {Carbotte}\ \emph {et~al.}(1999)\citenamefont
  {Carbotte}, \citenamefont {Schachinger},\ and\ \citenamefont
  {Basov}}]{carbotte:1999}%
  \BibitemOpen
  \bibfield  {author} {\bibinfo {author} {\bibfnamefont {J.~P.}\ \bibnamefont
  {Carbotte}}, \bibinfo {author} {\bibfnamefont {E.}~\bibnamefont
  {Schachinger}}, \ and\ \bibinfo {author} {\bibfnamefont {D.~N.}\ \bibnamefont
  {Basov}},\ }\href@noop {} {\bibfield  {journal} {\bibinfo  {journal} {Nature
  (London)}\ }\textbf {\bibinfo {volume} {401}},\ \bibinfo {pages} {354}
  (\bibinfo {year} {1999})}\BibitemShut {NoStop}%
\bibitem [{\citenamefont {Dordevic}\ \emph {et~al.}(2001)\citenamefont
  {Dordevic}, \citenamefont {Basov}, \citenamefont {Dilley}, \citenamefont
  {Bauer},\ and\ \citenamefont {Maple}}]{dordevic:2001}%
  \BibitemOpen
  \bibfield  {author} {\bibinfo {author} {\bibfnamefont {S.~V.}\ \bibnamefont
  {Dordevic}}, \bibinfo {author} {\bibfnamefont {D.~N.}\ \bibnamefont {Basov}},
  \bibinfo {author} {\bibfnamefont {N.~R.}\ \bibnamefont {Dilley}}, \bibinfo
  {author} {\bibfnamefont {E.~D.}\ \bibnamefont {Bauer}}, \ and\ \bibinfo
  {author} {\bibfnamefont {M.~B.}\ \bibnamefont {Maple}},\ }\href@noop {}
  {\bibfield  {journal} {\bibinfo  {journal} {Phys. Rev. Lett.}\ }\textbf
  {\bibinfo {volume} {86}},\ \bibinfo {pages} {684} (\bibinfo {year}
  {2001})}\BibitemShut {NoStop}%
\bibitem [{\citenamefont {Tran}\ \emph {et~al.}(2002)\citenamefont {Tran},
  \citenamefont {Donovan},\ and\ \citenamefont {Gruner}}]{tran:2002}%
  \BibitemOpen
  \bibfield  {author} {\bibinfo {author} {\bibfnamefont {P.}~\bibnamefont
  {Tran}}, \bibinfo {author} {\bibfnamefont {S.}~\bibnamefont {Donovan}}, \
  and\ \bibinfo {author} {\bibfnamefont {G.}~\bibnamefont {Gruner}},\
  }\href@noop {} {\bibfield  {journal} {\bibinfo  {journal} {Phys. Rev. B}\
  }\textbf {\bibinfo {volume} {65}},\ \bibinfo {pages} {205102} (\bibinfo
  {year} {2002})}\BibitemShut {NoStop}%
\bibitem [{\citenamefont {Dordevic}\ \emph {et~al.}(2005)\citenamefont
  {Dordevic}, \citenamefont {Homes}, \citenamefont {Tu}, \citenamefont {Valla},
  \citenamefont {Strongin}, \citenamefont {Johnson}, \citenamefont {Gu},\ and\
  \citenamefont {Basov}}]{dordevic:2005}%
  \BibitemOpen
  \bibfield  {author} {\bibinfo {author} {\bibfnamefont {S.~V.}\ \bibnamefont
  {Dordevic}}, \bibinfo {author} {\bibfnamefont {C.~C.}\ \bibnamefont {Homes}},
  \bibinfo {author} {\bibfnamefont {J.~J.}\ \bibnamefont {Tu}}, \bibinfo
  {author} {\bibfnamefont {T.}~\bibnamefont {Valla}}, \bibinfo {author}
  {\bibfnamefont {M.}~\bibnamefont {Strongin}}, \bibinfo {author}
  {\bibfnamefont {P.~D.}\ \bibnamefont {Johnson}}, \bibinfo {author}
  {\bibfnamefont {G.~D.}\ \bibnamefont {Gu}}, \ and\ \bibinfo {author}
  {\bibfnamefont {D.~N.}\ \bibnamefont {Basov}},\ }\href@noop {} {\bibfield
  {journal} {\bibinfo  {journal} {Phys. Rev. B}\ }\textbf {\bibinfo {volume}
  {71}},\ \bibinfo {pages} {104529} (\bibinfo {year} {2005})}\BibitemShut
  {NoStop}%
\bibitem [{\citenamefont {van Heumen}\ \emph {et~al.}(2009)\citenamefont {van
  Heumen}, \citenamefont {Muhlethaler}, \citenamefont {Kuzmenko}, \citenamefont
  {Eisaki}, \citenamefont {Meevasana}, \citenamefont {Greven},\ and\
  \citenamefont {van derMarel}}]{heumen:2009}%
  \BibitemOpen
  \bibfield  {author} {\bibinfo {author} {\bibfnamefont {E.}~\bibnamefont {van
  Heumen}}, \bibinfo {author} {\bibfnamefont {E.}~\bibnamefont {Muhlethaler}},
  \bibinfo {author} {\bibfnamefont {A.~B.}\ \bibnamefont {Kuzmenko}}, \bibinfo
  {author} {\bibfnamefont {H.}~\bibnamefont {Eisaki}}, \bibinfo {author}
  {\bibfnamefont {W.}~\bibnamefont {Meevasana}}, \bibinfo {author}
  {\bibfnamefont {M.}~\bibnamefont {Greven}}, \ and\ \bibinfo {author}
  {\bibfnamefont {D.}~\bibnamefont {van derMarel}},\ }\href@noop {} {\bibfield
  {journal} {\bibinfo  {journal} {Phys. Rev. B}\ }\textbf {\bibinfo {volume}
  {79}},\ \bibinfo {pages} {184512} (\bibinfo {year} {2009})}\BibitemShut
  {NoStop}%
\bibitem [{\citenamefont {Hwang}(2016{\natexlab{a}})}]{hwang:2016}%
  \BibitemOpen
  \bibfield  {author} {\bibinfo {author} {\bibfnamefont {J.}~\bibnamefont
  {Hwang}},\ }\href@noop {} {\bibfield  {journal} {\bibinfo  {journal} {J.
  Phys.: Condens. Matter}\ }\textbf {\bibinfo {volume} {28}},\ \bibinfo {pages}
  {125702} (\bibinfo {year} {2016}{\natexlab{a}})}\BibitemShut {NoStop}%
\bibitem [{\citenamefont {Collins}\ \emph {et~al.}(1989)\citenamefont
  {Collins}, \citenamefont {Schlesinger}, \citenamefont {Holtzberg},
  \citenamefont {Chaudhari},\ and\ \citenamefont {Feild}}]{collins:1989}%
  \BibitemOpen
  \bibfield  {author} {\bibinfo {author} {\bibfnamefont {R.~T.}\ \bibnamefont
  {Collins}}, \bibinfo {author} {\bibfnamefont {Z.}~\bibnamefont
  {Schlesinger}}, \bibinfo {author} {\bibfnamefont {F.}~\bibnamefont
  {Holtzberg}}, \bibinfo {author} {\bibfnamefont {P.}~\bibnamefont
  {Chaudhari}}, \ and\ \bibinfo {author} {\bibfnamefont {C.}~\bibnamefont
  {Feild}},\ }\href@noop {} {\bibfield  {journal} {\bibinfo  {journal} {Phys.
  Rev. B}\ }\textbf {\bibinfo {volume} {39}},\ \bibinfo {pages} {6571}
  (\bibinfo {year} {1989})}\BibitemShut {NoStop}%
\bibitem [{\citenamefont {Carbotte}(1990)}]{carbotte:1990}%
  \BibitemOpen
  \bibfield  {author} {\bibinfo {author} {\bibfnamefont {J.~P.}\ \bibnamefont
  {Carbotte}},\ }\href@noop {} {\bibfield  {journal} {\bibinfo  {journal} {Rev.
  Mod. Phys.}\ }\textbf {\bibinfo {volume} {62}},\ \bibinfo {pages} {1027}
  (\bibinfo {year} {1990})}\BibitemShut {NoStop}%
\bibitem [{\citenamefont {Schachinger}\ and\ \citenamefont
  {Carbotte}(2000)}]{schachinger:2000}%
  \BibitemOpen
  \bibfield  {author} {\bibinfo {author} {\bibfnamefont {E.}~\bibnamefont
  {Schachinger}}\ and\ \bibinfo {author} {\bibfnamefont {J.~P.}\ \bibnamefont
  {Carbotte}},\ }\href@noop {} {\bibfield  {journal} {\bibinfo  {journal}
  {Phys. Rev. B}\ }\textbf {\bibinfo {volume} {62}},\ \bibinfo {pages} {9054}
  (\bibinfo {year} {2000})}\BibitemShut {NoStop}%
\bibitem [{\citenamefont {Hwang}\ \emph {et~al.}(2006)\citenamefont {Hwang},
  \citenamefont {Yang}, \citenamefont {Timusk}, \citenamefont {Sharapov},
  \citenamefont {Carbotte}, \citenamefont {Bonn}, \citenamefont {Liang},\ and\
  \citenamefont {Hardy}}]{hwang:2006}%
  \BibitemOpen
  \bibfield  {author} {\bibinfo {author} {\bibfnamefont {J.}~\bibnamefont
  {Hwang}}, \bibinfo {author} {\bibfnamefont {J.}~\bibnamefont {Yang}},
  \bibinfo {author} {\bibfnamefont {T.}~\bibnamefont {Timusk}}, \bibinfo
  {author} {\bibfnamefont {S.~G.}\ \bibnamefont {Sharapov}}, \bibinfo {author}
  {\bibfnamefont {J.~P.}\ \bibnamefont {Carbotte}}, \bibinfo {author}
  {\bibfnamefont {D.~A.}\ \bibnamefont {Bonn}}, \bibinfo {author}
  {\bibfnamefont {R.}~\bibnamefont {Liang}}, \ and\ \bibinfo {author}
  {\bibfnamefont {W.~N.}\ \bibnamefont {Hardy}},\ }\href@noop {} {\bibfield
  {journal} {\bibinfo  {journal} {Phys. Rev. B}\ }\textbf {\bibinfo {volume}
  {73}},\ \bibinfo {pages} {014508} (\bibinfo {year} {2006})}\BibitemShut
  {NoStop}%
\bibitem [{\citenamefont {Hwang}\ \emph
  {et~al.}(2007{\natexlab{b}})\citenamefont {Hwang}, \citenamefont {Timusk},
  \citenamefont {Schachinger},\ and\ \citenamefont {Carbotte}}]{hwang:2007}%
  \BibitemOpen
  \bibfield  {author} {\bibinfo {author} {\bibfnamefont {J.}~\bibnamefont
  {Hwang}}, \bibinfo {author} {\bibfnamefont {T.}~\bibnamefont {Timusk}},
  \bibinfo {author} {\bibfnamefont {E.}~\bibnamefont {Schachinger}}, \ and\
  \bibinfo {author} {\bibfnamefont {J.~P.}\ \bibnamefont {Carbotte}},\
  }\href@noop {} {\bibfield  {journal} {\bibinfo  {journal} {Phys. Rev. B}\
  }\textbf {\bibinfo {volume} {75}},\ \bibinfo {pages} {144508} (\bibinfo
  {year} {2007}{\natexlab{b}})}\BibitemShut {NoStop}%
\bibitem [{\citenamefont {Schachinger}\ \emph {et~al.}(2006)\citenamefont
  {Schachinger}, \citenamefont {Neuber},\ and\ \citenamefont
  {Carbotte}}]{schachinger:2006}%
  \BibitemOpen
  \bibfield  {author} {\bibinfo {author} {\bibfnamefont {E.}~\bibnamefont
  {Schachinger}}, \bibinfo {author} {\bibfnamefont {D.}~\bibnamefont {Neuber}},
  \ and\ \bibinfo {author} {\bibfnamefont {J.~P.}\ \bibnamefont {Carbotte}},\
  }\href@noop {} {\bibfield  {journal} {\bibinfo  {journal} {Phys. Rev. B}\
  }\textbf {\bibinfo {volume} {73}},\ \bibinfo {pages} {184507} (\bibinfo
  {year} {2006})}\BibitemShut {NoStop}%
\bibitem [{\citenamefont {Hwang}(2015)}]{hwang:2015a}%
  \BibitemOpen
  \bibfield  {author} {\bibinfo {author} {\bibfnamefont {J.}~\bibnamefont
  {Hwang}},\ }\href@noop {} {\bibfield  {journal} {\bibinfo  {journal} {J.
  Phys.: Condens. Matter}\ }\textbf {\bibinfo {volume} {27}},\ \bibinfo {pages}
  {085701} (\bibinfo {year} {2015})}\BibitemShut {NoStop}%
\bibitem [{\citenamefont {Dordevic}\ \emph {et~al.}(2014)\citenamefont
  {Dordevic}, \citenamefont {van~der Marel},\ and\ \citenamefont
  {Homes}}]{dordevic:2014}%
  \BibitemOpen
  \bibfield  {author} {\bibinfo {author} {\bibfnamefont {S.~V.}\ \bibnamefont
  {Dordevic}}, \bibinfo {author} {\bibfnamefont {D.}~\bibnamefont {van~der
  Marel}}, \ and\ \bibinfo {author} {\bibfnamefont {C.~C.}\ \bibnamefont
  {Homes}},\ }\href@noop {} {\bibfield  {journal} {\bibinfo  {journal} {Phys.
  Rev. B}\ }\textbf {\bibinfo {volume} {90}},\ \bibinfo {pages} {174508}
  (\bibinfo {year} {2014})}\BibitemShut {NoStop}%
\bibitem [{\citenamefont {Carbotte}\ \emph {et~al.}(2011)\citenamefont
  {Carbotte}, \citenamefont {Timusk},\ and\ \citenamefont
  {Hwang}}]{carbotte:2011}%
  \BibitemOpen
  \bibfield  {author} {\bibinfo {author} {\bibfnamefont {J.~P.}\ \bibnamefont
  {Carbotte}}, \bibinfo {author} {\bibfnamefont {T.}~\bibnamefont {Timusk}}, \
  and\ \bibinfo {author} {\bibfnamefont {J.}~\bibnamefont {Hwang}},\
  }\href@noop {} {\bibfield  {journal} {\bibinfo  {journal} {Reports on
  Progress in Physics}\ }\textbf {\bibinfo {volume} {74}},\ \bibinfo {pages}
  {066501} (\bibinfo {year} {2011})}\BibitemShut {NoStop}%
\bibitem [{\citenamefont {Hwang}(2016{\natexlab{b}})}]{hwang:2016a}%
  \BibitemOpen
  \bibfield  {author} {\bibinfo {author} {\bibfnamefont {J.}~\bibnamefont
  {Hwang}},\ }\href@noop {} {\bibfield  {journal} {\bibinfo  {journal}
  {Scientific Reports}\ }\textbf {\bibinfo {volume} {6}},\ \bibinfo {pages}
  {23647} (\bibinfo {year} {2016}{\natexlab{b}})}\BibitemShut {NoStop}%
\bibitem [{\citenamefont {Hwang}(2018)}]{hwang:2018}%
  \BibitemOpen
  \bibfield  {author} {\bibinfo {author} {\bibfnamefont {J.}~\bibnamefont
  {Hwang}},\ }\href@noop {} {\bibfield  {journal} {\bibinfo  {journal} {J.
  Phys.: Condens. Matter}\ }\textbf {\bibinfo {volume} {30}},\ \bibinfo {pages}
  {405604} (\bibinfo {year} {2018})}\BibitemShut {NoStop}%
\bibitem [{\citenamefont {Allen}(1971)}]{allen:1971}%
  \BibitemOpen
  \bibfield  {author} {\bibinfo {author} {\bibfnamefont {P.~B.}\ \bibnamefont
  {Allen}},\ }\href@noop {} {\bibfield  {journal} {\bibinfo  {journal} {Phys.
  Rev. B}\ }\textbf {\bibinfo {volume} {3}},\ \bibinfo {pages} {305} (\bibinfo
  {year} {1971})}\BibitemShut {NoStop}%
\bibitem [{\citenamefont {Hwang}\ \emph {et~al.}(2008)\citenamefont {Hwang},
  \citenamefont {Yang}, \citenamefont {Carbotte},\ and\ \citenamefont
  {Timusk}}]{hwang:2008a}%
  \BibitemOpen
  \bibfield  {author} {\bibinfo {author} {\bibfnamefont {J.}~\bibnamefont
  {Hwang}}, \bibinfo {author} {\bibfnamefont {J.}~\bibnamefont {Yang}},
  \bibinfo {author} {\bibfnamefont {J.~P.}\ \bibnamefont {Carbotte}}, \ and\
  \bibinfo {author} {\bibfnamefont {T.}~\bibnamefont {Timusk}},\ }\href@noop {}
  {\bibfield  {journal} {\bibinfo  {journal} {J. Phys. Condens. Matter}\
  }\textbf {\bibinfo {volume} {20}},\ \bibinfo {pages} {295215} (\bibinfo
  {year} {2008})}\BibitemShut {NoStop}%
\bibitem [{\citenamefont {Nussenzveig}(1972)}]{nussenzveig:1972}%
  \BibitemOpen
  \bibfield  {author} {\bibinfo {author} {\bibfnamefont {H.~M.}\ \bibnamefont
  {Nussenzveig}},\ }\href@noop {} {\emph {\bibinfo {title} {Causality and
  Dispersion Relations}}}\ (\bibinfo  {publisher} {Accademic Press,
  Massachusetts, USA},\ \bibinfo {year} {1972})\ \bibinfo {note} {(Note: Vol.
  95, p.30)}\BibitemShut {NoStop}%
\bibitem [{\citenamefont {Wooten}(1972)}]{wooten}%
  \BibitemOpen
  \bibfield  {author} {\bibinfo {author} {\bibfnamefont {F.}~\bibnamefont
  {Wooten}},\ }\href@noop {} {\emph {\bibinfo {title} {Optical Properties of
  Solids}}}\ (\bibinfo  {publisher} {Academic, New York},\ \bibinfo {year}
  {1972})\ \bibinfo {note} {(Note: Key material on page 176)}\BibitemShut
  {NoStop}%
\bibitem [{\citenamefont {Glover}\ and\ \citenamefont
  {Tinkham}(1956)}]{glover:1956}%
  \BibitemOpen
  \bibfield  {author} {\bibinfo {author} {\bibfnamefont {R.~E.}\ \bibnamefont
  {Glover}}\ and\ \bibinfo {author} {\bibfnamefont {M.}~\bibnamefont
  {Tinkham}},\ }\href@noop {} {\bibfield  {journal} {\bibinfo  {journal} {Phys.
  Rev.}\ }\textbf {\bibinfo {volume} {104}},\ \bibinfo {pages} {844} (\bibinfo
  {year} {1956})}\BibitemShut {NoStop}%
\bibitem [{\citenamefont {Ferrell}\ and\ \citenamefont
  {Glover}(1958)}]{ferrell:1958}%
  \BibitemOpen
  \bibfield  {author} {\bibinfo {author} {\bibfnamefont {R.~A.}\ \bibnamefont
  {Ferrell}}\ and\ \bibinfo {author} {\bibfnamefont {R.~E.}\ \bibnamefont
  {Glover}},\ }\href@noop {} {\bibfield  {journal} {\bibinfo  {journal} {Phys.
  Rev.}\ }\textbf {\bibinfo {volume} {109}},\ \bibinfo {pages} {1398} (\bibinfo
  {year} {1958})}\BibitemShut {NoStop}%
\bibitem [{\citenamefont {Hwang}\ \emph
  {et~al.}(2007{\natexlab{c}})\citenamefont {Hwang}, \citenamefont {Timusk},\
  and\ \citenamefont {Gu}}]{hwang:2007a}%
  \BibitemOpen
  \bibfield  {author} {\bibinfo {author} {\bibfnamefont {J.}~\bibnamefont
  {Hwang}}, \bibinfo {author} {\bibfnamefont {T.}~\bibnamefont {Timusk}}, \
  and\ \bibinfo {author} {\bibfnamefont {G.~D.}\ \bibnamefont {Gu}},\
  }\href@noop {} {\bibfield  {journal} {\bibinfo  {journal} {J. Phys.: Condens.
  Matter}\ }\textbf {\bibinfo {volume} {19}},\ \bibinfo {pages} {125208}
  (\bibinfo {year} {2007}{\natexlab{c}})}\BibitemShut {NoStop}%
\bibitem [{\citenamefont {Millis}\ \emph {et~al.}(1990)\citenamefont {Millis},
  \citenamefont {Monien},\ and\ \citenamefont {Pines}}]{millis:1990}%
  \BibitemOpen
  \bibfield  {author} {\bibinfo {author} {\bibfnamefont {A.~J.}\ \bibnamefont
  {Millis}}, \bibinfo {author} {\bibfnamefont {H.}~\bibnamefont {Monien}}, \
  and\ \bibinfo {author} {\bibfnamefont {D.}~\bibnamefont {Pines}},\
  }\href@noop {} {\bibfield  {journal} {\bibinfo  {journal} {Phys. Rev. B}\
  }\textbf {\bibinfo {volume} {42}},\ \bibinfo {pages} {167} (\bibinfo {year}
  {1990})}\BibitemShut {NoStop}%
\bibitem [{\citenamefont {Basov}\ and\ \citenamefont
  {Timusk}(2005)}]{basov:2005}%
  \BibitemOpen
  \bibfield  {author} {\bibinfo {author} {\bibfnamefont {D.~N.}\ \bibnamefont
  {Basov}}\ and\ \bibinfo {author} {\bibfnamefont {T.}~\bibnamefont {Timusk}},\
  }\href@noop {} {\bibfield  {journal} {\bibinfo  {journal} {Rev. Mod. Phys}\
  }\textbf {\bibinfo {volume} {77}},\ \bibinfo {pages} {721} (\bibinfo {year}
  {2005})}\BibitemShut {NoStop}%
\bibitem [{\citenamefont {Rossat-Mignod}\ \emph {et~al.}(1991)\citenamefont
  {Rossat-Mignod}, \citenamefont {Regnault}, \citenamefont {Vettier},
  \citenamefont {Bourges}, \citenamefont {Burlet}, \citenamefont {Bossy},
  \citenamefont {Henry},\ and\ \citenamefont {Lapertot}}]{rossat:1991}%
  \BibitemOpen
  \bibfield  {author} {\bibinfo {author} {\bibfnamefont {J.}~\bibnamefont
  {Rossat-Mignod}}, \bibinfo {author} {\bibfnamefont {L.~P.}\ \bibnamefont
  {Regnault}}, \bibinfo {author} {\bibfnamefont {C.}~\bibnamefont {Vettier}},
  \bibinfo {author} {\bibfnamefont {P.}~\bibnamefont {Bourges}}, \bibinfo
  {author} {\bibfnamefont {P.}~\bibnamefont {Burlet}}, \bibinfo {author}
  {\bibfnamefont {J.}~\bibnamefont {Bossy}}, \bibinfo {author} {\bibfnamefont
  {J.~Y.}\ \bibnamefont {Henry}}, \ and\ \bibinfo {author} {\bibfnamefont
  {G.}~\bibnamefont {Lapertot}},\ }\href@noop {} {\bibfield  {journal}
  {\bibinfo  {journal} {Physica C}\ }\textbf {\bibinfo {volume} {185-189}},\
  \bibinfo {pages} {86} (\bibinfo {year} {1991})}\BibitemShut {NoStop}%
\bibitem [{\citenamefont {Dai}\ \emph {et~al.}(1999)\citenamefont {Dai},
  \citenamefont {Mook}, \citenamefont {Hayden}, \citenamefont {Aeppli},
  \citenamefont {Perring}, \citenamefont {Hunt},\ and\ \citenamefont
  {Doguan}}]{dai:1999}%
  \BibitemOpen
  \bibfield  {author} {\bibinfo {author} {\bibfnamefont {P.}~\bibnamefont
  {Dai}}, \bibinfo {author} {\bibfnamefont {H.~A.}\ \bibnamefont {Mook}},
  \bibinfo {author} {\bibfnamefont {S.~M.}\ \bibnamefont {Hayden}}, \bibinfo
  {author} {\bibfnamefont {G.}~\bibnamefont {Aeppli}}, \bibinfo {author}
  {\bibfnamefont {T.~G.}\ \bibnamefont {Perring}}, \bibinfo {author}
  {\bibfnamefont {R.~D.}\ \bibnamefont {Hunt}}, \ and\ \bibinfo {author}
  {\bibfnamefont {F.}~\bibnamefont {Doguan}},\ }\href@noop {} {\bibfield
  {journal} {\bibinfo  {journal} {Science}\ }\textbf {\bibinfo {volume}
  {284}},\ \bibinfo {pages} {1344} (\bibinfo {year} {1999})}\BibitemShut
  {NoStop}%
\bibitem [{\citenamefont {Fong}\ \emph {et~al.}(2000)\citenamefont {Fong},
  \citenamefont {Bourges}, \citenamefont {Sidis}, \citenamefont {Regnault},
  \citenamefont {Bossy}, \citenamefont {Ivanov}, \citenamefont {Milius},
  \citenamefont {Aksay},\ and\ \citenamefont {Keimer}}]{fong:2000}%
  \BibitemOpen
  \bibfield  {author} {\bibinfo {author} {\bibfnamefont {H.~F.}\ \bibnamefont
  {Fong}}, \bibinfo {author} {\bibfnamefont {P.}~\bibnamefont {Bourges}},
  \bibinfo {author} {\bibfnamefont {Y.}~\bibnamefont {Sidis}}, \bibinfo
  {author} {\bibfnamefont {L.~P.}\ \bibnamefont {Regnault}}, \bibinfo {author}
  {\bibfnamefont {J.}~\bibnamefont {Bossy}}, \bibinfo {author} {\bibfnamefont
  {A.}~\bibnamefont {Ivanov}}, \bibinfo {author} {\bibfnamefont {D.~L.}\
  \bibnamefont {Milius}}, \bibinfo {author} {\bibfnamefont {I.~A.}\
  \bibnamefont {Aksay}}, \ and\ \bibinfo {author} {\bibfnamefont
  {B.}~\bibnamefont {Keimer}},\ }\href@noop {} {\bibfield  {journal} {\bibinfo
  {journal} {Phys. Rev. B}\ }\textbf {\bibinfo {volume} {61}},\ \bibinfo
  {pages} {14773} (\bibinfo {year} {2000})}\BibitemShut {NoStop}%
\bibitem [{\citenamefont {Stock}\ \emph {et~al.}(2004)\citenamefont {Stock},
  \citenamefont {Buyers}, \citenamefont {Liang}, \citenamefont {Peets},
  \citenamefont {Tun}, \citenamefont {Bonn}, \citenamefont {Hardy}, ,\ and\
  \citenamefont {Birgeneau}}]{stock:2004}%
  \BibitemOpen
  \bibfield  {author} {\bibinfo {author} {\bibfnamefont {C.}~\bibnamefont
  {Stock}}, \bibinfo {author} {\bibfnamefont {W.~J.~L.}\ \bibnamefont
  {Buyers}}, \bibinfo {author} {\bibfnamefont {R.}~\bibnamefont {Liang}},
  \bibinfo {author} {\bibfnamefont {D.}~\bibnamefont {Peets}}, \bibinfo
  {author} {\bibfnamefont {Z.}~\bibnamefont {Tun}}, \bibinfo {author}
  {\bibfnamefont {D.}~\bibnamefont {Bonn}}, \bibinfo {author} {\bibfnamefont
  {W.~N.}\ \bibnamefont {Hardy}}, , \ and\ \bibinfo {author} {\bibfnamefont
  {R.~J.}\ \bibnamefont {Birgeneau}},\ }\href@noop {} {\bibfield  {journal}
  {\bibinfo  {journal} {Phys. Rev. B}\ }\textbf {\bibinfo {volume} {69}},\
  \bibinfo {pages} {014502} (\bibinfo {year} {2004})}\BibitemShut {NoStop}%
\bibitem [{\citenamefont {Zasadzinski}\ \emph {et~al.}(2006)\citenamefont
  {Zasadzinski}, \citenamefont {Ozyuzer}, \citenamefont {Coffey}, \citenamefont
  {Gray}, \citenamefont {Hinks},\ and\ \citenamefont
  {Kendziora}}]{zasadzinski:2006}%
  \BibitemOpen
  \bibfield  {author} {\bibinfo {author} {\bibfnamefont {J.~F.}\ \bibnamefont
  {Zasadzinski}}, \bibinfo {author} {\bibfnamefont {L.}~\bibnamefont
  {Ozyuzer}}, \bibinfo {author} {\bibfnamefont {L.}~\bibnamefont {Coffey}},
  \bibinfo {author} {\bibfnamefont {K.~E.}\ \bibnamefont {Gray}}, \bibinfo
  {author} {\bibfnamefont {D.~G.}\ \bibnamefont {Hinks}}, \ and\ \bibinfo
  {author} {\bibfnamefont {C.}~\bibnamefont {Kendziora}},\ }\href@noop {}
  {\bibfield  {journal} {\bibinfo  {journal} {Phys. Rev. Lett.}\ }\textbf
  {\bibinfo {volume} {96}},\ \bibinfo {pages} {017004} (\bibinfo {year}
  {2006})}\BibitemShut {NoStop}%
\bibitem [{\citenamefont {Johnson}\ \emph {et~al.}(2001)\citenamefont
  {Johnson}, \citenamefont {Valla}, \citenamefont {Fedorov}, \citenamefont
  {Yusof}, \citenamefont {Wells}, \citenamefont {Li}, \citenamefont
  {Moodenbaugh}, \citenamefont {Gu}, \citenamefont {Koshizuka}, \citenamefont
  {Kendziora}, \citenamefont {Jian},\ and\ \citenamefont
  {Hinks}}]{johnson:2001}%
  \BibitemOpen
  \bibfield  {author} {\bibinfo {author} {\bibfnamefont {P.~D.}\ \bibnamefont
  {Johnson}}, \bibinfo {author} {\bibfnamefont {T.}~\bibnamefont {Valla}},
  \bibinfo {author} {\bibfnamefont {A.~V.}\ \bibnamefont {Fedorov}}, \bibinfo
  {author} {\bibfnamefont {Z.}~\bibnamefont {Yusof}}, \bibinfo {author}
  {\bibfnamefont {B.~O.}\ \bibnamefont {Wells}}, \bibinfo {author}
  {\bibfnamefont {Q.}~\bibnamefont {Li}}, \bibinfo {author} {\bibfnamefont
  {A.~R.}\ \bibnamefont {Moodenbaugh}}, \bibinfo {author} {\bibfnamefont
  {G.~D.}\ \bibnamefont {Gu}}, \bibinfo {author} {\bibfnamefont
  {N.}~\bibnamefont {Koshizuka}}, \bibinfo {author} {\bibfnamefont
  {C.}~\bibnamefont {Kendziora}}, \bibinfo {author} {\bibfnamefont
  {S.}~\bibnamefont {Jian}}, \ and\ \bibinfo {author} {\bibfnamefont {D.~G.}\
  \bibnamefont {Hinks}},\ }\href@noop {} {\bibfield  {journal} {\bibinfo
  {journal} {Phys. Rev. Lett.}\ }\textbf {\bibinfo {volume} {87}},\ \bibinfo
  {pages} {177007} (\bibinfo {year} {2001})}\BibitemShut {NoStop}%
\bibitem [{\citenamefont {Valla}\ \emph {et~al.}(2007)\citenamefont {Valla},
  \citenamefont {Kidd}, \citenamefont {Yin}, \citenamefont {Gu}, \citenamefont
  {Johnson}, \citenamefont {Pan},\ and\ \citenamefont {Fedorov}}]{valla:2007}%
  \BibitemOpen
  \bibfield  {author} {\bibinfo {author} {\bibfnamefont {T.}~\bibnamefont
  {Valla}}, \bibinfo {author} {\bibfnamefont {T.~E.}\ \bibnamefont {Kidd}},
  \bibinfo {author} {\bibfnamefont {W.-G.}\ \bibnamefont {Yin}}, \bibinfo
  {author} {\bibfnamefont {G.~D.}\ \bibnamefont {Gu}}, \bibinfo {author}
  {\bibfnamefont {P.~D.}\ \bibnamefont {Johnson}}, \bibinfo {author}
  {\bibfnamefont {Z.-H.}\ \bibnamefont {Pan}}, \ and\ \bibinfo {author}
  {\bibfnamefont {A.~V.}\ \bibnamefont {Fedorov}},\ }\href@noop {} {\bibfield
  {journal} {\bibinfo  {journal} {Phys. Rev. Lett.}\ }\textbf {\bibinfo
  {volume} {98}},\ \bibinfo {pages} {167003} (\bibinfo {year}
  {2007})}\BibitemShut {NoStop}%
\bibitem [{\citenamefont {Zhang}\ \emph {et~al.}(2008)\citenamefont {Zhang},
  \citenamefont {Liu}, \citenamefont {Zhao}, \citenamefont {Liu}, \citenamefont
  {Meng}, \citenamefont {Dong}, \citenamefont {Lu}, \citenamefont {Wen},
  \citenamefont {Xu}, \citenamefont {Gu}, \citenamefont {Sasagawa},
  \citenamefont {Wang}, \citenamefont {Zhu}, \citenamefont {Zhang},
  \citenamefont {Zhou}, \citenamefont {Wang}, \citenamefont {Zhao},
  \citenamefont {Chen}, \citenamefont {Xu},\ and\ \citenamefont
  {Zhou}}]{zhang:2008}%
  \BibitemOpen
  \bibfield  {author} {\bibinfo {author} {\bibfnamefont {W.}~\bibnamefont
  {Zhang}}, \bibinfo {author} {\bibfnamefont {G.}~\bibnamefont {Liu}}, \bibinfo
  {author} {\bibfnamefont {L.}~\bibnamefont {Zhao}}, \bibinfo {author}
  {\bibfnamefont {H.}~\bibnamefont {Liu}}, \bibinfo {author} {\bibfnamefont
  {J.}~\bibnamefont {Meng}}, \bibinfo {author} {\bibfnamefont {X.}~\bibnamefont
  {Dong}}, \bibinfo {author} {\bibfnamefont {W.}~\bibnamefont {Lu}}, \bibinfo
  {author} {\bibfnamefont {J.~S.}\ \bibnamefont {Wen}}, \bibinfo {author}
  {\bibfnamefont {Z.~J.}\ \bibnamefont {Xu}}, \bibinfo {author} {\bibfnamefont
  {G.~D.}\ \bibnamefont {Gu}}, \bibinfo {author} {\bibfnamefont
  {T.}~\bibnamefont {Sasagawa}}, \bibinfo {author} {\bibfnamefont
  {G.}~\bibnamefont {Wang}}, \bibinfo {author} {\bibfnamefont {Y.}~\bibnamefont
  {Zhu}}, \bibinfo {author} {\bibfnamefont {H.}~\bibnamefont {Zhang}}, \bibinfo
  {author} {\bibfnamefont {Y.}~\bibnamefont {Zhou}}, \bibinfo {author}
  {\bibfnamefont {X.}~\bibnamefont {Wang}}, \bibinfo {author} {\bibfnamefont
  {Z.}~\bibnamefont {Zhao}}, \bibinfo {author} {\bibfnamefont {C.}~\bibnamefont
  {Chen}}, \bibinfo {author} {\bibfnamefont {Z.}~\bibnamefont {Xu}}, \ and\
  \bibinfo {author} {\bibfnamefont {X.~J.}\ \bibnamefont {Zhou}},\ }\href@noop
  {} {\bibfield  {journal} {\bibinfo  {journal} {Phys. Rev. Lett.}\ }\textbf
  {\bibinfo {volume} {100}},\ \bibinfo {pages} {107002} (\bibinfo {year}
  {2008})}\BibitemShut {NoStop}%
\bibitem [{\citenamefont {Hwang}(2013)}]{hwang:2013a}%
  \BibitemOpen
  \bibfield  {author} {\bibinfo {author} {\bibfnamefont {J.}~\bibnamefont
  {Hwang}},\ }\href@noop {} {\bibfield  {journal} {\bibinfo  {journal} {J.
  Phys. Condens. Matter}\ }\textbf {\bibinfo {volume} {25}},\ \bibinfo {pages}
  {295701} (\bibinfo {year} {2013})}\BibitemShut {NoStop}%
\bibitem [{\citenamefont {Schachinger}\ and\ \citenamefont
  {Carbotte}(2017)}]{schachinger:2017}%
  \BibitemOpen
  \bibfield  {author} {\bibinfo {author} {\bibfnamefont {E.}~\bibnamefont
  {Schachinger}}\ and\ \bibinfo {author} {\bibfnamefont {J.~P.}\ \bibnamefont
  {Carbotte}},\ }\href@noop {} {\bibfield  {journal} {\bibinfo  {journal}
  {Phys. Rev. B}\ }\textbf {\bibinfo {volume} {95}},\ \bibinfo {pages} {144516}
  (\bibinfo {year} {2017})}\BibitemShut {NoStop}%
\end{thebibliography}%

\end{document}